# Downgrade Attack on TrustZone

Yue Chen[1], Yulong Zhang[2], Zhi Wang[1], Tao Wei[2]

[1]Florida State University

[2]Baidu X-Lab

## ABSTRACT

Security-critical tasks require proper isolation from untrusted software. Chip manufacturers design and include trusted execution environments (TEEs) in their processors to secure these tasks. The integrity and security of the software in the trusted environment depend on the verification process of the system.

We find a form of attack that can be performed on the current implementations of the widely deployed ARM TrustZone technology. The attack exploits the fact that the trustlet (TA) or TrustZone OS loading verification procedure may use the same verification key and may lack proper rollback prevention across versions. If an exploit works on an out-of-date version, but the vulnerability is patched on the latest version, an attacker can still use the same exploit to compromise the latest system by downgrading the software to an older and exploitable version.

We did experiments on popular devices on the market including those from Google, Samsung and Huawei, and found that all of them have the risk of being attacked. Also, we show a real-world example to exploit Qualcomm's QSEE.

In addition, in order to find out which device images share the same verification key, pattern matching schemes for different vendors are analyzed and summarized.

## 1. BACKGROUND

To enhance the security of today's computer systems, hardware manufacturers have introduced a new security mechanism called trusted execution environment (TEE), like ARM's TrustZone (TZ) [1], which has a widespread deployment in the digital world.

Basically, it has two separate worlds: one is called *normal world* and another one is *secure world*. Each world has its own operating system (OS) and user applications as shown in Figure 1. In practice, the normal world contains *untrusted* software, and the secure world contains *trusted* software. By design, the normal world is isolated from the secure world by hardware-enforced mechanisms.

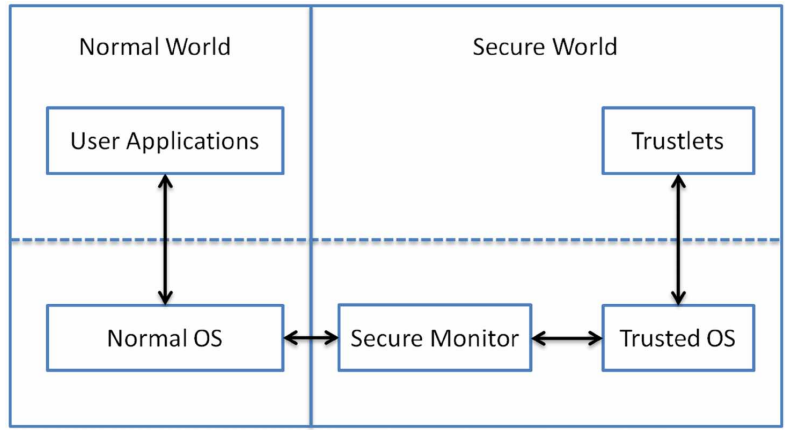

Figure 1: Overview of the typical TEE architecture

Simply put, the user applications and normal OS in the normal world are the traditional ones, while the ones in secure world have dedicated usages (e.g., digital rights management, authentication, etc.). We call the OS (privileged) in the secure world trusted OS, and the user-space applications (unprivileged) are called trustlets (also called trusted applications (TAs)). The two worlds communicate through the secure monitor.

When the trusted OS loads a trustlet from its unprivileged mode, it firstly checks its signature to see if it is signed by the right party and if the software is modified. This integrity check aims at removing the risk of loading tampered trustlets.

By design, starting from the hardware (e.g., eFuse/qFuse or ROM), a chain of trust is formed. This chain includes the bootloaders, the trusted OS, and derived certificates or keys. This step-by-step verification procedure ensures the integrity of the secure world. Potentially, the security and integrity of the whole system relies on this kind of verification.

## 2.  ATTACK DESCRIPTION

### 2.1 The Attack

When a trustlet is being loaded, its cryptographic digital signature will be validated. Usually this signature is computed using a private key to encrypt the hash of the trustlet (or certain parts of the trustlet). Only the correct public key can decrypt it, and then the trustlet can be verified. For security purposes, the key pairs are based on the chain of trust, originating from the hardware.

Ideally, if the system is updated, the older software like trustlets cannot be loaded into the newer system. However, the trustlets can be replaced with their corresponding older versions. If the older version has a vulnerability and the newer version is patched, the attacker can still exploit the patched version by replacing it with an older one. We call it downgrade attack or rollback attack.

A successful exploit first needs to have the root privilege of the device (e.g., exploit another vulnerability), and then use this issue combined with other vulnerabilities to exploit the device, potentially compromising the TrustZone/TEE (even its kernel).

This vulnerability potentially impacts all the devices that are on the current market, though we do not have a thorough test and experiment.

In section 2.3, we will discuss the *impact* of downgrade attack on more components of TEE, rather than only on trustlet loading.

## 2.2 Experiment

We did experiments on several phones from major vendors, including Google Nexus 5 and Nexus 6, Samsung Galaxy S7, Huawei Mate 9. The results show that this attack works on all of them. We also have read the documents and code of the open-source OP-TEE [2], and found that it potentially has the same problem, though we did not do experiments on its platform.

To reproduce the procedure, the steps are as follows:

1. Root the device.

2. Remount the file system that contains the trustlets (e.g., "mount -o rw,remount /system").

3. Replace the current trustlets with the corresponding (vulnerable) ones from an older-version image.

4. Use the device as normal.

To test if the trustlets can still work correctly, we could use apps that can communicate with the corresponding truetlets to check its functionality. Or we can write one by ourselves (e.g., `QSEECom_start_app` for QSEE). Another approach is to read TrustZone logs. The logs stay in different places for different models, here are some possible locations:

dmesg

/sys/kernel/debug/

/d/

/dev/hisi_teelog     (Huawei's devices)

Also note that the trustlets are in different formats as well as file extensions and paths on different vendors' platforms. Note that for the same phone model, there may exist *different* patterns for different versions or builds. Some of the parameters in our experimentation are listed for further reference:

Nexus 5 and Nexus 6:

/system/vendor/firmware/

The trustlets have a set of files. For example, for widevine:

widevine.b00

widevine.b01

widevine.b02

widevine.b03

widevine.mdt

Samsung Galaxy S7:

/system/app/mcRegistry/

The trustlets are named as [UUID].tlbin

(In some older versions, the path is /data/app/mcRegistry/)

Huawei Mate 9:

/vendor/bin/

The trustlets are named as [UUID].sec

Other paths also contain some *.sec files, like

/product/bin/

/sbin/

On Nexus 6, we run a **real-world exploit** [3,4] on Nexus 6 and successfully use the downgrade attack to exploit a QSEE privilege escalation vulnerability.

For Nexus 6, this vulnerability could be exploited in version LMY48M, but not N6F26Y. We replace N6F26Y's widevine trustlets with the ones from LMY48M. Then the exploit (CVE-2015-6639) works.

## 2.3 Attack on TrustZone OS, Bootloader, and More

The secure boot procedure has a chain of trust. In its boot sequence, each software image to be executed is authenticated by software that was previously verified. This design prevents unauthorized or tampered code from being run. Different devices could have different boot stages, including different bootloaders. For example, for Qualcomm's MSM8960 chipset, `SBL1` loads `SBL2`, and `SBL2` loads `tz` and `SBL3`. Here the `SBL` stands for secondary bootloader.

Similar to the loading verification of trustlets, the TrustZone OS also needs loading verification. Hence, it is under the haze of downgrade attack, too. In our experiments with Google Nexus 5, the TrustZone OS, even the bootloader, can be downgraded without failure. Since the principle and experimentation are similar to trustlet verification, we do not dig into the details here.

We believe that this problem exists on more components in the chain of trust, potentially affecting the fundamental security of the whole system.

## 2.4 Solution

To defeat this attack, the obvious solution is to use different key pairs for different versions. However, due to the complexity of the chain of trust, device distribution and compatibility issues, this method is relatively difficult to implement for the whole ecosystem especially for old devices.

Another reasonable and ideal solution is to use version control for rollback prevention. To make it easy, one can increase the version number once finding vulnerabilities in its older version. In reality, some TEE vendors already have this mechanism implemented (see Section 3). However, based on the responses from phone vendors, the solution is also not so practical. First, like the first solution, it may take too many resources to completely fix this issue especially for old devices on the current market. Second, the successful exploit depends on other vulnerabilities. If all other vulnerabilities are fixed or not related to this issue, the devices cannot be exploited. Third, taking into user experience into consideration, this issue somewhat can be regarded as a feature. For example, end users may want to downgrade certain parts of the system to meet their own needs.

Based on the responses from vendors up to now, Samsung is supporting the fix since Galaxy S8/S8+ devices.

# 3.   COMPATIBILITY MATCHING

If one needs to write a scanner to see if two image versions can load their trustlets mutually, or to cluster images into such groups, here we provide a pattern matching analysis on the TEE implementations of different device vendors. To ensure the compatibility of the trustlets across versions, they may need the API compatibility with the TEE OS, and may need to pass the version check if there is any. Qualcomm does have the version control parameter `SW_ID` [5], but it is not properly configured, as most of the images have the value of `0`. So the most important is the signature verification. This procedure requires the correct public key or certificate to verify the trustlet. If it fails, the trustlet cannot be loaded. Until now in our experiments, we find that the keys are the determining factor for load verificaiton. So in the next part we will focus on this factor and describe our findings on different platforms.

Through reverse engineering and analysis, we try to scan the TEE OS images to see which trustlet/image pairs can be mutually replaceable.

All the following analyses are based on our reverse engineering, observations and guesswork. The correctness and generality are *not* guaranteed.

### 3.1 Google Nexus

```
1330h: 65 63 78 D2 09 40 05 5E BA B9 56 13 C5 BB FF C1   e c x Ò . @ . ^ º ¹ V . Å » ÿ Á
1340h: 52 2F 86 5F ED 10 1A 1D 30 82 04 C4 30 82 03 AC   R / † _ í . . . 0 . . Ä 0 . . ¬
1350h: A0 03 02 01 02 02 03 00 97 66 30 0D 06 09 2A 86   . . . . . . . . — f 0 . . . * †
1360h: 48 86 F7 0D 01 01 0B 05 00 30 7C 31 0B 30 09 06   H † ÷ . . . . . 0 | 1 . 0 . .
1370h: 03 55 04 06 13 02 55 53 31 0B 30 09 06 03 55 04   . U . . . . U S 1 . 0 . . . U .
1380h: 08 13 02 43 41 31 12 30 10 06 03 55 04 07 13 09   . . . C A 1 . 0 . . . U . . . .
1390h: 53 61 6E 20 44 69 65 67 6F 31 1A 30 18 06 03 55   S a n   D i e g o 1 . 0 . . . U
13A0h: 04 0B 13 11 43 44 4D 41 20 54 65 63 68 6E 6F 6C   . . . . C D M A   T e c h n o l
13B0h: 6F 67 69 65 73 31 0D 30 0B 06 03 55 04 0A 13 04   o g i e s 1 . 0 . . . U . . . .
13C0h: 4E 6F 6E 65 31 21 30 1F 06 03 55 04 03 13 18 47   N o n e 1 ! 0 . . . U . . . . G
13D0h: 65 6E 65 72 61 74 65 64 20 41 74 74 65 73 74 61   e n e r a t e d   A t t e s t a
13E0h: 74 69 6F 6E 20 43 41 30 1E 17 0D 31 37 30 32 32   t i o n   C A 0 . . . 1 7 0 2 2
13F0h: 32 30 39 34 32 31 35 5A 17 0D 33 37 30 32 31 37   2 0 9 4 2 1 5 Z . . 3 7 0 2 1 7
1400h: 30 39 34 32 31 35 5A 30 82 01 51 30 0B 30 09 06   0 9 4 2 1 5 Z 0 . . Q 0 . 0 . .
1410h: 03 55 04 06 13 02 55 53 31 0B 30 09 06 03 55 04   . U . . . U S 1 . 0 . . . U .
1420h: 03 13 2A 51 75 61 6C 63 6F 6D 6D 20 50 6C 61 74   . . * Q u a l c o m m   P l a t
1430h: 66 6F 72 6D 20 53 69 67 6E 69 6E 67 20 41 70 70   f o r m   S i g n i n g   A p p
1440h: 6C 69 63 61 74 69 6F 6E 20 55 73 65 72 31 12 30   l i c a t i o n   U s e r 1 . 0
1450h: 10 06 03 55 04 07 13 09 53 61 6E 20 44 69 65 67   . . . U . . . . S a n   D i e g
1460h: 6F 31 0D 30 0B 06 03 55 04 0A 13 04 41 53 49 43   o 1 . 0 . . . U . . . . A S I C
1470h: 31 13 30 11 06 03 55 04 08 13 0A 43 61 6C 69 66   1 . 0 . . . U . . . C a l i f
1480h: 6F 72 6E 69 61 31 17 30 15 06 03 55 04 0B 14 0E   o r n i a 1 . 0 . . . U . . . .
1490h: 30 34 20 30 30 30 20 4F 45 4D 5F 49 44 31 1C   0 4   0 0 0   O E M _ I D 1 .
14A0h: 30 1A 06 03 55 04 0B 14 13 30 35 20 30 30 30 30   0 . . . U . . . 0 5   0 0 0 0
14B0h: 30 32 34 38 20 53 57 5F 53 49 5A 45 31 19 30 17   0 2 4 8   S W _ S I Z E 1 . 0 .
14C0h: 06 03 55 04 0B 14 10 30 36 20 30 30 20 4D   . . U . . . 0 6   0 0   M
14D0h: 4F 44 45 4C 5F 49 44 31 17 30 15 06 03 55 04 0B   O D E L _ I D 1 . 0 . . . U . .
14E0h: 13 0E 30 37 20 30 30 30 31 20 53 48 41 32 35 36   . . 0 7   0 0 0 1   S H A 2 5 6
```

Figure 2: The certificate pattern in the "tz" image of Google Nexus 6

First we take a look at Google Nexus 6 (Nexus 5 is very similar). The TrustZone image can be found at /dev/block/platform/msm_sdcc.1/by-name/. Its name is "tz". Another one called "tzBackup" ("tzb" in Nexus 5) is its identical backup [6,7].

The certificates are embedded in the image. There are three certificates (could be two in other devices) that form a certificate chain. We open the image in a hex editor. As shown in Figure 2, if the hexadecimal pattern is

**30 82 [XX] [XX] 30 82 ...**

we can say that it is a beginning of a certificate with DER (Distinguished Encoding Rules) encoding. Note that **30 82** are **0x30** and **0x82** in hex, and the two bytes in between can be random.

Once finding the beginning of a certificate, we can save the raw data from the hex editor, and the certificate can be decoded using openssl:

```
openssl x509 -in cert.der -inform der -text
```

Here the saved data length could be the distance starting from this pattern to the next same pattern. The actual length does not matter since openssl will automatically identify the certificate and discard the following useless part. Here the certificates follow the X.509 v3 format.

Note that in Figure 2, there exists another (third) "**30 82**" at the line beginning at address **1400h**. This one is not the described pattern and does not affect the decoding, therefore we can ignore it. Once getting the certificates, we can compare them against the ones from another image to see if their trustlets can be mutually replaced.

## 3.2 Samsung

Samsung has two TrustZone vendors. One is Qualcomm, usually for the phones in North America; another one is Trustonic (formerly MobiCore), for the rest of the world. We analyze them separately in our study.

### 3.2.1 Qualcomm:

The Qualcomm TruztZone image is a standalone one, called "tz.mbn". As shown in Figure 3, the pattern is very similar to the one in Google Nexus 6. This is because their vendors are the same, Qualcomm. And its TEE is called QSEE (Qualcomm's Secure Execution Environment). So we can use the same pattern to extract the certificates and compare them from different image versions.

Figure 3: The certificate pattern in the "tz.mbn" image of Samsung S7 (USA version)

Through reversing engineering, we also found two other RSA public keys with the exponent of 65537, embedded in the TrustZone image. We guess they are used for other purposes.

### 3.2.2 Trustonic/Mobicore:

Trustonic (formerly Mobicore) is another TrustZone vendor for Samsung's mobile devices. Its TEE implementation is embedded in the image "sboot.bin".

With the investigation into its images, we found a reasonable pattern as follows:

**`00 01 00 00 [256 bytes] 01 00 00 00 03 ...`**

The example is shown in Figure 4. What needs to be compared is the 256-byte blob between "**`00 01 00 00`**" and "**`01 00 00 00 03`**". A standalone 256-byte blob cannot form an RSA-2048 public key or a standard certificate. However, since the RSA public exponent is usually set as 65537 or 3, which can be easily embedded elsewhere, we guess this 256-byte (2048-bit) blob is the modulus of an RSA public key, which is enough for the comparison purpose.

Figure 4: The public key pattern in the "sboot.bin" image of Samsung S7 (UAE version)

## 3.3 Huawei

Huawei's TrustZone image is named as "TEEOS.img", and can be extracted from its update package "UPDATE.APP".

We did reverse engineering analysis on the raw image with a series of steps, including symbol lookup table recovery, segment fix, etc. The result shows that it does not have the needed public key or certificate embedded in the firmware. We found that the public key used for Trusted Application (TA) signature verification derives from the hardware.

As shown in Figure 5, the blob of **`0`**'s starting at **`seg1:0x0017D73C`** is used to hold the derived public key. This area will be filled with the key during runtime. So we are unable to extract it statically. This screenshot is taken from IDA Pro, on one version of Huawei Mate 9 Pro. During disassembly, we create a new segment so the addresses are different from the original ones.

On the TEE user-space part, interestingly, Huawei's TAs (trustlets) are encrypted. Thus analysis on them needs decryption first.

Figure 5: The unloaded public key region in the "TEEOS.img" image of Huawei Mate 9 Pro

## 3.4 OP-TEE

Since the OP-TEE platform is open-source, it is relatively easy to analyze its logic and trustlet loading verification mechanism. So we omit its analysis here.

# 4. RELATED WORK

TEE's strong-isolation capability and its security-sensitivity make it an interesting topic for security researchers and industrial security practitioners. Several vulnerabilities have been discovered to break certain TEE implementations [4,8,9,10,11,12]. A set of security systems leveraging TEE are also proposed [13,14]. Other related works include TrustZone virtualization [15] and analysis on vendor implementations [16,17].

# 5. CONCLUSION

In this work, we give an overview of the TEE design, describe the downgrade attack on TEE, and show how to reproduce the experiment results in detail. In addition, the analyses on the compatibility matching schemes are given for further reference.

# REFERENCES

[1] ARM. ARM TrustZone. http://www.arm.com/products/processors/technologies/trustzone/index.php.

[2] OP-TEE. https://github.com/OP-TEE.

[3] CVE-2015-6639. https://nvd.nist.gov/vuln/detail/CVE-2015-6639

[4] CVE-2015-6639 Exploit. https://github.com/laginimaineb/cve-2015-6639.

[5] Qualcomm. Secure Boot and Image Authentication Technical Overview. https://www.qualcomm.com/documents/secure-boot-and-image-authentication-technical-overview.

[6] laginimaineb. Bits, Please!. http://bits-please.blogspot.com/2015/08/exploring-qualcomms-trustzone.html

[7] laginimaineb. Bits, Please!. http://bits-please.blogspot.com/2016/04/exploring-qualcomms-secure-execution.html

[8] Rosenberg, D. QSEE TrustZone Kernel Integer Overflow Vulnerability. Black Hat USA 2014.

[9] Shen, D. Exploiting Trustzone on Android. Black Hat USA 2015.

[10] Zhang, Y., Chen, Z., Xue, H. and Wei, T. Fingerprints On Mobile Devices: Abusing and Leaking. Black Hat USA 2015.

[11] Machiry, A., Gustafson, E., Spensky, C., Salls, C., Stephens, N., Wang, R., Bianchi, A., Choe, Y.R., Kruegel, C. and Vigna, G. BOOMERANG: Exploiting the Semantic Gap in Trusted Execution Environments. NDSS 2017.

[12] Lee, J., Jang, J., Jang, Y., Kwak, N., Choi, Y., Choi, C., Kim, T., Peinado, M. and Kang, B.B. Hacking in Darkness: Return-oriented Programming against Secure Enclaves. USENIX Security 2017.

[13] Azab, A.M., Ning, P., Shah, J., Chen, Q., Bhutkar, R., Ganesh, G., Ma, J. and Shen, W.

Hypervision Across Worlds: Real-time Kernel Protection from the ARM TrustZone Secure World. CCS 2014.

[14] Guan, L., Liu, P., Xing, X., Ge, X., Zhang, S., Yu, M. and Jaeger, T. TrustShadow: Secure Execution of Unmodified Applications with ARM TrustZone. MobiSys 2017.

[15] Hua, Z., Gu, J., Xia, Y., Chen, H., Zang, B. and Guan, H. vTZ: Virtualizing ARM TrustZone. USENIX Security 2017.

[16] Kanonov, U. and Wool, A. Secure Containers in Android: the Samsung KNOX Case Study. SPSM 2016.

[17] Atamli-Reineh, A., Borgaonkar, R., Balisane, R.A., Petracca, G. and Martin, A. Analysis of Trusted Execution Environment usage in Samsung KNOX. SysTEX 2016.